\documentclass[twocolumn,amsmath,amssymb,showpacs,superscriptaddress]{revtex4}
\usepackage{epsf}
\usepackage{epsfig}
\usepackage{graphicx}
\usepackage{dcolumn}
\usepackage{bm}
\usepackage{amsmath}
\usepackage{amsfonts}
\usepackage{amssymb}
\input epsf

\begin{document}

\title{Inelastic Quantum Transport and Peierls-like Mechanism in Carbon Nanotubes}
\author{Luis E. F. \surname{Foa Torres}}
\affiliation{CEA/DSM/DRFMC/SPSMS/GT, 17 avenue des Martyrs, 38054 Grenoble, France}
\affiliation{CEA/DRT/LETI/DIHS/LMNO, 17 avenue des Martyrs, 38054 Grenoble, France}
\author{Stephan Roche}
\affiliation{CEA/DSM/DRFMC/SPSMS/GT, 17 avenue des Martyrs, 38054 Grenoble, France}

\date{\today}
\begin{abstract}
We report on a theoretical study of inelastic quantum transport in $(3m,0)$ carbon nanotubes. By using a many-body description of the electron-phonon interaction in Fock space, a novel mechanism involving optical phonon emission (absorption) is shown to induce an unprecedented energy gap opening at half the phonon energy, $\hbar\omega_{0}/2$, above (below) the charge neutrality point. This mechanism, which is prevented by Pauli blocking at low bias voltages, is activated at
bias voltages in the order of $\hbar\omega_{0}$.

\end{abstract}
\maketitle

\preprint{APS/123-QED}




%
%
%
%
%
%

%
%
%
%
%
%
%

%
%
%
%
%
Since their discovery, carbon nanotubes (CNTs) have attracted much attention
due to their outstanding mechanical and electrical properties \cite{SaitoBook}%
. Depending on their helicity, CNTs exhibit either semiconducting or metallic
behavior and for low resistance contacts, ballistic transport is observed in
the low bias regime. At higher bias, several experimental works report on
current saturation, attributed to inelastic (inter-band) backscattering by
optical phonons \cite{DekkerPRL2000}. Electron-phonon (e-ph) interaction thus
severely limits the exceptional properties of CNTs as ballistic conductors.
Aimed to the explanation of the experimental data
\cite{DekkerPRL2000,DaiHotPhonons,JaveyPRL2004}, several theoretical studies
of electronic transport in CNTs in presence of e-ph interaction were performed
\cite{KurodaLeburtonPRL2005,SvizhenkoPRB2005,Cuniberti,SRoche-Phonons,MauriHotPhononsFGR,MauriHotPhononsBoltzmann}%
. They include simulations based on the Boltzmann equation
\cite{KurodaLeburtonPRL2005,MauriHotPhononsBoltzmann}, the Fermi Golden Rule
\cite{MauriHotPhononsFGR} and the use of a diagonal self-consistent Born approximation
\cite{SvizhenkoPRB2005}. Besides, in order to explain the quantitative
disagreement between theoretical and experimental estimations of inelastic
mean free paths, a hot phonons scenario was recently proposed
\cite{MauriHotPhononsFGR}.

On the other hand, electron-phonon coupling in low-dimensional systems can lead to more subtle effects and important corrections to both phonon and electronic bandstructures.
Fundamental examples are the Kohn anomaly \cite{MauriKohnAnomalies} and the
Peierls transition \cite{MintmirePRL1992,BlasePeierlsdistorsion}. Connected to
the latter, and by using density functional theory, Dubay and coworkers found
a softening of the mode with $A_{1}$ symmetry in metallic CNTs, not sufficient
however, to produce a static lattice distortion \cite{DubayKresse}.
Notwithstanding, as explored within the linear response regime, the
corresponding phonon-induced (time-dependent) electronic bandstructure changes
result in strong modifications of the Kubo conductance \cite{SRoche-Phonons}.
This points towards the importance for a quantum mechanical treatment of
inelastic transport whenever high energy (optical) modes are activated.

In this Letter, by using a full quantum description of the joined processes of
tunneling and phonon-assisted transport, the coupling between electrons and
optical $A_{1}(L)$ phonons in $(N=3m,0)$ zig-zag tubes is shown to result in an
energy-gap opening at $\hbar\omega_{0}/2$ above (below) the charge neutrality
point (CNP), owing to phonon emission (absorption). This novel many-body mechanism is
activated when driving the system out of equilibrium.

\begin{figure}[ptb]
%
%
%
%
%
\begin{center}
\includegraphics[width=0.44\textwidth]{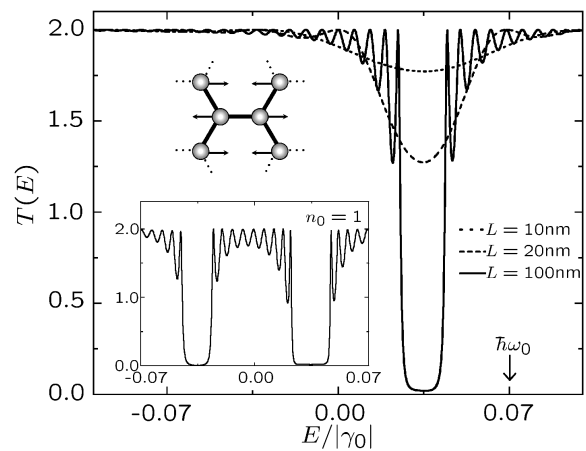}
\vspace*{-0.5em}
\end{center}
\caption{Total transmission probability as a function of
the energy of the incident electrons for $n_{0}=0$. Inset: Same information for
$n_{0}=1$. The results correspond to a $(24,0)$ tube where interaction with an
$A_{1}(L)$ mode inducing displacements along the axis direction (drawing) is
included. Note that a total energy $\varepsilon=\hbar\omega_{0}/2$ corresponds to an
electronic kinetic energy $E=\hbar\omega_{0}/2$ for $n=0$ and $-\hbar
\omega_{0}/2$ for $n=1$.}%
\label{fig-Ts}%
\end{figure}

For simplicity, we consider an infinite CNT and allow the electrons to
interact with phonons only in a central part of length $L$. The Hamiltonian of
the system is written as a sum of an electronic part, a phonon part and the
e-ph interaction term. The electronic part is described by an effective $\pi
$-orbitals model $H_{e}=-\gamma_{0}^{{}} \sum_{\left\langle i,j\right\rangle
}\left[  c_{i}^{+}c_{j}^{{}}+h.c.\right]  $, where $c_{i}^{+}$ and $c_{i}^{{}%
}$ are the creation and annihilation operators for electrons at site $i$,
$\gamma_{0}=2.77$eV is the transfer integral restricted to nearest neighbors
$\pi$-orbitals \cite{SaitoBook}. The phonon term is given by $H_{ph}%
=\hbar\omega_{0}b_{{}}^{+}b_{{}}^{{}},$ with $b_{{}}^{+}$ and $b_{{}}^{{}}$
the phonon operators. The remaining contribution is given by a
Su-Schreiffer-Heeger Hamiltonian \cite{SSH}, describing the e-ph coupling to a
non-local vibrational eigenmode. This is found by assuming a phonon modulation
of the hopping matrix elements \cite{SSH,MahanPRB2003}, keeping only the
linear corrections to the atomic displacements from equilibrium, i.e.
$\gamma_{i,j}=\gamma_{0}+\alpha\widehat{\delta}_{i,j}\cdot\delta
\vec{Q}_{i,j}$, where $\widehat{\delta}_{i,j}$ is a unit vector in
the bond direction, $\delta\vec{Q}_{i,j}$ is the relative
displacement of the neighboring carbon atoms, and $\alpha$ is the e-ph
coupling strength defined as the derivative of $\gamma_{0}$ with respect to
the change in the bond length. Further quantization of the atomic
displacements gives the e-ph interaction term:%
\begin{equation}
H_{e\text{-}ph}=\sum_{\left\langle i,j\right\rangle _{vib}}\left[
\gamma_{i,j}^{e\text{-}ph}\,c_{i}^{+}c_{j}^{{}}(b_{{}}^{+}+b_{{}}^{{}%
})+h.c.\right]  ,
\end{equation}
where the e-ph matrix elements can be written as $\gamma_{i,j}^{e\text{-}%
ph}=\alpha\sqrt{\hbar/\left(  2m\omega_{0}\right)  }\widehat{\delta}%
_{i,j}\cdot(\widehat{e}_{i}-\widehat{e}_{j})$, $\widehat{e}_{i}$ is the phonon
mode eigenvector which gives the atomic displacements. The summation in the
right hand side corresponds to nearest neighbors within the vibrating region
of the CNT. For the phonon mode considered here ($A_{1}(L)$), A and B-type
atoms move out of phase in the direction parallel to the tube axis (see scheme
in Fig. \ref{fig-Ts}).

To compute inelastic quantum transport, we use the approach introduced in
Refs. \cite{BJPAnda1994,BoncaTrugman1995}, which has been applied to a variety
of problems including vibration-assisted tunneling in STM experiments
\cite{cit-MingoPRL2000}, transport through molecules
\cite{Ness-e-ph,cit-Kirczenow-e-ph} and resonant tunneling in double barrier
heterostructures \cite{cit-PRBFock2001}. This scheme allows an exact
non-perturbative treatment of the problem of one electron transport in the
presence of e-ph interaction. This is achieved by an exact mapping of the
many-body problem into a single particle problem in a higher dimensional space
(the e-ph Fock space), each considered phonon mode adding one extra dimension
to the problem. The key idea behind this approach can be visualized after
re-writing the interacting Hamiltonian in an appropriate basis for the e-ph Fock
space (one electron plus phonons). In this equivalent multichannel one-body problem, the asymptotic states
(in the non-interacting leads) include both the electronic and the vibrational
degrees of freedom and the total energy is fully conserved. When considering a
single phonon mode, these asymptotic channels can be labeled by using two
indexes: $(X,n)$, being $X=L,R$ the index associated to the corresponding
electrode (left or right) and $n$ the number of phonon excitations in the
system. Accordingly, the transmission $T_{(X,n)\rightarrow(Y,m)}^{{}}$ and
reflection probabilities $R_{(X,n)\rightarrow(Y,m)}^{{}}$ between the
different channels are computed by using standard Green's functions techniques
\cite{cit-PRBFock2001}. Using these probabilities as inputs, the
non-equilibrium electron distributions in the leads (at finite temperature and
bias voltage) are evaluated self-consistently using the procedure developed in
Ref. \cite{cit-Kirczenow-e-ph}. The electronic current through the device can
be obtained from these self-consistent distributions which take into account 
the Pauli exclusion principle for the different competing processes.

Let us first explicitly establish how the Fock states are connected by the
Hamiltonian and simplify the problem by using a mode decomposition. The
matrix elements of the e-ph Hamiltonian are essentially given by the projection of
the bond direction on the relative displacements of the atoms from their
equilibrium positions, $\widehat{\delta}_{i,j}\cdot(\widehat{e}_{i}%
-\widehat{e}_{j})$. It can be easily proven that the e-ph matrix elements
$\gamma_{i,j}^{e\text{-}ph}$ have values $\gamma
_{0}^{e\text{-}ph}=-2\alpha\sqrt{\hbar/\left(  2m\omega_{0}\right)  }$ for
bonds $(i,j)$ that are parallel to the tube axis, and $-\gamma_{0}%
^{e\text{-}ph}\cos(\pi/3)$ for bonds $(i,j)$ that are tilted with respect to
the tube axis.

For the phonon mode considered here, instead of solving the Hamiltonian in
real space, one uses a mode space approach \cite{cit-Mingo2001,SvizhenkoPRB2005}.
The idea is to resort to a unitary transformation that diagonalizes the
electronic Hamiltonian for each layer perpendicular to the tube axis. The
resulting eigenstates will correspond to different circumferential modes that
can be used to build an alternative basis for the description of transport
through the nanotube. In absence of static disorder, the electronic
Hamiltonian $H_{e}$ does not couple the different modes, and the different
subbands correspond to linear chains with alternating hoppings $\gamma_{0}$
and $\gamma_{q}=2\gamma_{0}\cos(q\pi/N)$ ($q=0,1...,N-1$) with dispersion
relations: $\varepsilon^{(0)}(k)=\pm\sqrt{\gamma_{0}^{2}+\gamma_{q}%
^{2}+2\gamma_{0}\gamma_{q}\cos(3ka_{cc}/2)}$. When $N$ is an integer multiple
of three, the metallic subbands correspond to $q=N/3,2N/3$. Furthermore, since
the e-ph Hamiltonian $H_{e\text{-}ph}$ does not couple different
circumferential modes neither, one is left with $N$ independent problems. Each
of them corresponds to a one dimensional mode space lattice that, when coupled
to phonons, gives a two-dimensional problem in Fock space.

In Fig. \ref{fig-Ts}, one reports the total transmission probability
$T(E)=\sum_{n}T_{(L,n_{0})\rightarrow(R,n)}(E)$ as a function of the
incident's electron energy $E$, when there are initially no phonon
excitations in the system, $n_{0}=0$. All the energies are expressed in units
of $\gamma_{0}$, the value of $\hbar\omega_{0}$ is taken as $0.07\gamma_{0}$
\cite{DubayKresse} and $\alpha=\alpha_{0}\simeq7\mathrm{eV}/\AA$ is estimated from
\cite{PorezagPRB1995}. For short nanotube lengths ($L=10$nm), the main feature
is the occurence of a transmission dip centered at $E=\hbar\omega_{0}/2$ above
the CNP, that progressively deepens with CNT length, to reach a full gap at
$L=100$nm, and with a width of approximately $3\gamma_{0}^{e\text{-}ph}$.
Apart from this fundamental gap, the effect of e-ph interaction remains small. The transmission remains mostly elastic in all the energy range shown in the plot. However, the reduction
of the transmission probability in the gap region corresponds to a
complementary increase in the \textit{inelastic backscattering by phonon
emission}.

Figure \ref{fig-Ts}(inset) shows the transmission probability in the situation
where one phonon is already available for scattering before any additional
charge enters into the sample. In this case, in addition to the expected
enlargement of the gap at $E=\hbar\omega_{0}/2$ by a factor $\sqrt{2}$ due to
stimulated phonon emission, another gap develops at $E=-\hbar\omega_{0}/2$.
This transmission gap is complemented by an increase in the \textit{inelastic
backscattering by phonon absorption}.

\begin{figure}[ptb]
\begin{center}
\includegraphics[width=0.44\textwidth]{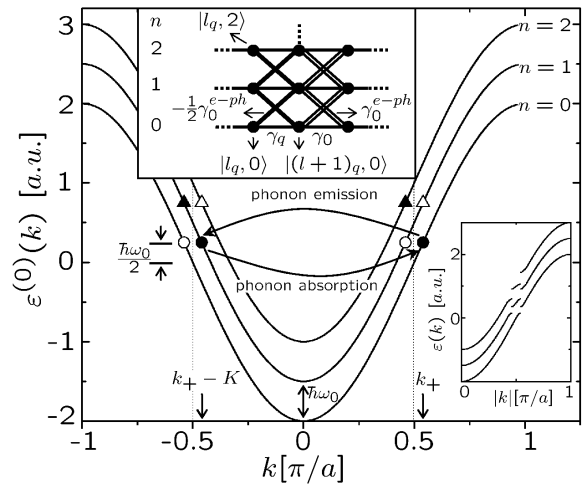}
\vspace*{-0.5em}
\end{center}
\caption{Scheme for the unperturbed ($H_{e-ph}=0$) dispersion relations
corresponding to $n=0,1,2$ for the mode $q=N/3$. The
states marked with identical symbols correspond to degenerate states of the
coupled e-ph system. These degeneracies are lifted by the e-ph interaction
leading to the opening of energy gaps at $\hbar\omega_{0}/2$ above (below) the
band center (see right inset). Upper inset: Representation of the
many-body Hamiltonian in Fock space for a circumferential mode $q$ (note the two-layer periodicity).
Solid circles represent states in Fock space while the lines are off-diagonal
matrix elements.}%
\label{fig-schematics}%
\end{figure}

To provide a clear physical picture of such phenomena, let us consider that
the sample, i.e. the region where the e-ph interaction is allowed, extends to
infinity. The Fock space for the coupled e-ph system can then be expanded in
terms of the basis states $\left\{  \left|  l_{q},n\right\rangle =\left|
l_{q}\right\rangle \otimes\left|  n\right\rangle \right\}  $, where $\left|
l_{q}\right\rangle $ is the q-th circumferential mode localized in the $l-$th
layer and $\left|  n\right\rangle $ corresponds to the state with $n$ phonons
in the system. Close to the CNP, only two circumferential modes participate in conduction ($q=N/3,2N/3$). The
e-ph Hamiltonian for one of those modes is represented in Fig.
\ref{fig-schematics} (upper inset). Alternatively, the Fock space for an
individual mode can be expanded by using plane waves instead of localized
states for the electronic part. This leads to basis states of the form
$\left|  k,n\right\rangle =\left|  k\right\rangle \otimes\left|
n\right\rangle $ where $\left|  k\right\rangle $ is a plane wave in mode space
with wave-vector $k$ along the axis direction.

Let us consider the circumferential mode with $q=N/3$, though a similar reasoning holds for $q=2N/3$.
In absence of e-ph coupling, one gets disconnected chains associated to the different values of $n$, each one with hoppings $\gamma_q=\gamma_0$ and unperturbed dispersion relations
$\varepsilon_{n}^{(0)}=\varepsilon_{n}^{(0)}(k)$ (see Fig.
\ref{fig-schematics}). Of particular interest are the states $\left|
k_{+}=\pi/2a+\left|  \delta k\right|  ,0\right\rangle $ and $\left|
k_{+}-K,1\right\rangle $ (black circles), where $a=$ $3a_{cc}/4$ and $K=\pi
/a$. They have the same total energy $\varepsilon_{0}^{(0)}(k_{+}%
)=\varepsilon_{1}^{(0)}(k_{+}-K)=\hbar\omega_{0}/2$. When the
e-ph interaction is switched on, the lattice period of the many-body Hamiltonian is doubled because of the spatial periodicy of $H_{e\text{-}ph}$. The new lattice vector is $K=\pi/a$ instead of $2\pi/a$. 
\textit{The mentioned Fock states are now mixed
by} $H_{e\text{-}ph}$, i.e.
$\left\langle k_{+}-K,1\right|  H_{e\text{-}ph}\left|  k_{+},0\right\rangle
=\Delta\gamma^{e\text{-}ph}\equiv\left(  3/2\right)  \gamma_{0}^{e\text{-}%
ph}\neq0$ \cite{footnote}. The degeneracy is thus lifted, giving rise to the opening of an
energy gap in $\varepsilon_{0}(k)$ and $\varepsilon_{1}(k)$ of width $2\left|
\Delta\gamma^{e\text{-}ph}\right|  $ around $\varepsilon=\hbar\omega_{0}/2$ 
(right inset in Fig. \ref{fig-schematics}). Accordingly, an incoming electron
with a wave-vector $k$ in the Fock state $\left|  k,n\right\rangle $ will
contain, as $k$ approaches $k_{+}$ (or $k_{+}-K$), an increasing
admixture of $\left|  k-K,n+1\right\rangle $ (or $\left|  k+K,n-1\right\rangle $)
leading to a Bragg-type of inelastic scattering. This mechanism is due to a
modification of the translational symmetry of the system driven by the e-ph interaction, with no static distortion of the lattice. A similar effect holds for the other states marked as
symbols of the same kind in Fig. \ref{fig-schematics} and for the sub-band with $q=2N/3$. This results in the transmission gaps shown in Fig. \ref{fig-Ts} which are in quantitative agreement with the prior analysis.

\begin{figure}[ptb]
%
%
%
%
%
%
%
%
%
\begin{center}
\includegraphics[width=0.44\textwidth]{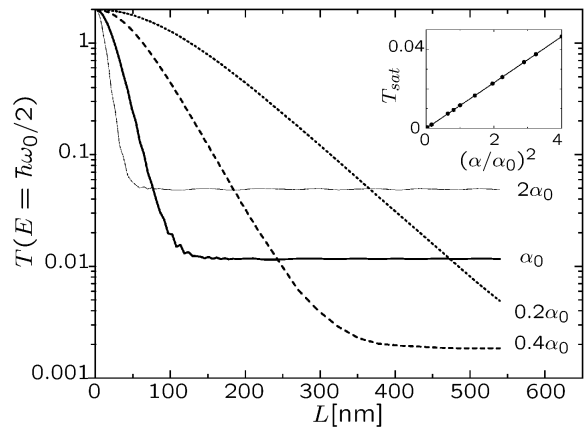}
\vspace*{-0.5em}
\end{center}
\caption{Total transmission probability at
$E=\hbar\omega_{0}/2$ as a function of the tube length for $n_{0}=0$, the other
parameters are the same as in Fig. \ref{fig-Ts}. The inset shows the
saturation value of the total transmission as a function of $\alpha^{2}$.}%
\label{fig-TsvsL}%
\end{figure}

Figure \ref{fig-TsvsL} shows the transmission minimum at the
center of the phonon emission gap $T(E=\hbar\omega_{0}/2)$ as a
function of the nanotube length. Different curves correspond to different
values of the parameter $\alpha$ entering into the e-ph matrix elements
$\gamma_{i,j}^{e\text{-}ph}$. From a simple argument, one can predict the
existence of a region where the transmission decays exponentially with the
tube length (tunneling through the gap), i.e. $T\propto\exp(-L/\xi)$, with a
decay length $\xi$ that is inversely proportional to the energy gap
$\Delta\gamma^{e\text{-}ph}$. Another interesting feature is the observed
saturation for longer tubes. The saturation value $T_{sat}$ as a function of
$\alpha^{2}$ (inset of Fig. \ref{fig-TsvsL}) is associated to the small inelastic component of the transmission.

\begin{figure}[ptb]
%
%
%
%
%
%
%
%
%
\begin{center}
\includegraphics[width=0.44\textwidth]{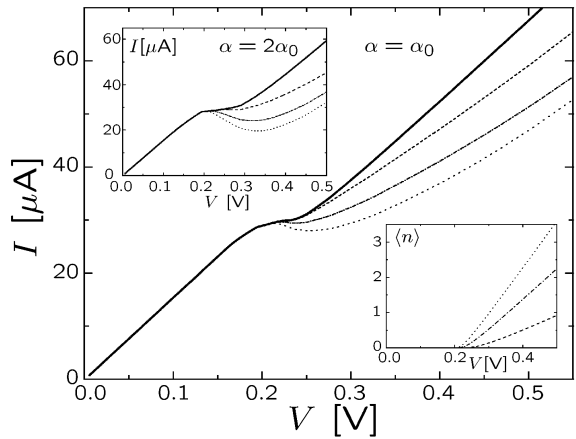}
\vspace*{-0.5em}
\end{center}
\caption{Current vs. bias voltage for a $(24,0)$ $150$
nm long tube for $\alpha=\alpha_{0}$ (main frame) and $\alpha=2\alpha_{0}$ (upper inset). 
Solid lines correspond to a fixed phonon population $n_{0}%
=0$. The dashed, dash-dotted and dotted lines correspond to the effective
thermal distributions described in the text. The mean number of phonons
corresponding to these curves are shown in the lower inset.}%
\label{fig-I-V}%
\end{figure}

It should be noted that the opening of the transmission gap will be prevented by Pauli blocking at low bias voltage. In order to activate it, the system has to be driven out of equilibrium by applying a
sufficiently high bias between the voltage probes. To explore the consequences
of this phenomenon in\ the current-voltage characteristics, the effect of the
bias voltage is introduced on $H_{e}$ by modifying the onsite energies of the
$\pi$-orbitals. The potential drop is assumed to be equally distributed at the
two contacts. The current-voltage curves calculated from this simple model for
$\alpha=\alpha_{0}$ \ and $\alpha=2\alpha_{0}$ and zero temperature are shown
in Fig. \ref{fig-I-V} (solid lines in the main frame and inset respectively).
The main feature that results from the gap opening is the onset of a current plateau observed at $V\sim\hbar\omega_{0}$ whose width scales linearly with $\alpha$. The subsequent linear increase in the current is due to the lack of other ingredients in our model such as electronic coupling with other phonon modes (such as those producing inter-band backscattering) or electrostatics effects.  Motivated by recent studies that suggest the possible build up of a strong out-of equilibrium
phonon population at high bias voltages
\cite{MauriHotPhononsFGR,DaiHotPhonons,MauriHotPhononsBoltzmann}, a natural
question is how such scenario would modify our results. The answer might be
provided by a self-consistent scheme to determine the out-of equilibrium
phonon population in the system, but this is beyond the scope of the present
study. Instead, a thermalized phonon population with an
effective temperature $k_{B}T_{eff}$ is assumed and $k_{B}T_{eff}(V)$ is
supposed to increase linearly for $eV\geq\hbar\omega_{0}$ \cite{MauriHotPhononsFGR,DaiHotPhonons,MauriHotPhononsBoltzmann}.
The dependence of the mean numbers of phonons with the bias voltage for
different slopes in $k_{B}T_{eff}(V)$ ($10$kK/V$,20$kK/V and $30$kK/V) are
shown in the lower inset. The corresponding currents for a $150$nm long tube
are shown with dashed, dash-dotted and dotted lines for $\alpha=\alpha_{0}$ 
(Fig. \ref{fig-I-V}) and $\alpha=2\alpha_{0}$ (upper inset). A substantial
decrease in the currents when the phonon population increases is observed.
Furthermore, the dotted lines reveal the appearance of a region of negative
differential resistance.

Semiconducting zig-zag CNTs with a bandgap smaller than $\hbar\omega_{0}$ are
also found to display the phonon-induced transmission gap (not shown here). On the other hand,
for CNTs other than zig-zag, this phenomenon will be driven by the
relevant optical phonon mode producing an instantaneous dimerization pattern
of appropriate period in the axis direction. In contrast to the Peierls transition, where a mean
field description is feasible \cite{SSH}, the proposed mechanism does not involve a
static lattice distortion and is activated at high bias voltage. 

This work was supported by the French Ministry of Research under grant 
RTB: PostCMOS moleculaire 200mm.

%
%
%
%
%


\begin{thebibliography}{99}
\bibitem{footnote}From the point of view of momentum conservation, this process can be seen as an Umklapp process.
\end{thebibliography}
\end{document}